\documentclass[10pt,letterpaper]{article}
\usepackage{opex3}

\usepackage[cmex10]{amsmath}
\usepackage{amssymb}
\usepackage{color}
\usepackage{subfigure}
\usepackage{array}

\newcommand{\myeqno}[1]{Eq.~(\ref{#1})}
\newcommand{\myfigno}[1]{Fig.~\ref{#1}}
\newcommand{\mytabno}[1]{Table~\ref{#1}}

\newcommand{\absval}[1]{\left\arrowvert{#1}\right\arrowvert}
\newcommand{\ssub}[1]{_{\mathrm{#1}}}

\newcommand{\mum}{\,\mu\mathrm{m}}
\newcommand{\mnm}{\,\mathrm{nm}}
\newcommand{\neff}{n\ssub{eff}}
\newcommand{\gc}{\gamma\ssub{c}}
\newcommand{\gs}{\gamma\ssub{s}}
\newcommand{\gcn}{\gamma_{\mathrm{c}n}}
\newcommand{\gsn}{\gamma_{\mathrm{s}n}}
\newcommand{\ac}{\alpha\ssub{c}}
\newcommand{\as}{\alpha\ssub{s}}
\newcommand{\xc}{\xi\ssub{c}}
\newcommand{\xs}{\xi\ssub{s}}
\newcommand{\acn}{\alpha_{\mathrm{c}n}}
\newcommand{\asn}{\alpha_{\mathrm{s}n}}
\newcommand{\Ks}{K\ssub{s}}
\newcommand{\Kc}{K\ssub{c}}
\newcommand{\Gs}{G\ssub{s}}
\newcommand{\Gc}{G\ssub{c}}
\newcommand{\Qc}{Q\ssub{c}}
\newcommand{\Qs}{Q\ssub{s}}
\newcommand{\Gcn}{G_{\mathrm{c}n}}
\newcommand{\Gsn}{G_{\mathrm{s}n}}

\newcommand{\epsf}{\epsilon\ssub{f}}
\newcommand{\epss}{\epsilon\ssub{s}}
\newcommand{\epsc}{\epsilon\ssub{c}}
\newcommand{\csch}{\mathrm{csch}}
\newcommand{\mtica}{Mathematica{\texttrademark}}

\begin{document}

\title{Solving dielectric and plasmonic waveguide dispersion relations on a pocket calculator}

\author{Rohan D.\ Kekatpure, Aaron C.\ Hryciw, Edward S.\ Barnard, and\\ Mark L.\ Brongersma}

\address{Geballe Laboratory for Advanced Materials, Stanford University, Stanford, CA, 94305}

\email{brongersma@stanford.edu} 



\begin{abstract}
We present a robust iterative technique for solving complex transcendental dispersion equations routinely encountered in integrated optics. Our method especially befits the multilayer dielectric and plasmonic waveguides forming the basis structures for a host of contemporary nanophotonic devices. The solution  algorithm ports seamlessly from the real to the complex domain---i.e., no extra complexity results when dealing with leaky structures or those with material/metal loss. Unlike several existing numerical approaches, our algorithm exhibits markedly-reduced sensitivity to the initial guess and allows for straightforward implementation on a pocket calculator.
\end{abstract}



\section{Introduction}\label{intro}
With an ever-improving ability to fabricate miniature functional components, the field of nanophotonics is in a phase of explosive growth. Not only are novel phenomena being routinely observed in wavelength- and sub-wavelength-scale components,~but these concepts are being rapidly translated into exotic devices for applications ranging from solar energy and information technology, to biology \cite{mark_book,rashid:mattoday,pala:advmat}. At the heart of this development is the ability of nanostructures to confine, guide, and scatter light in ways that are not achievable with bulk materials. It follows that understanding and improving the performance of these nanostructures requires a detailed knowledge of the electromagnetic modes they support.

 The allowed electromagnetic modes in microphotonic structures are determined by solving Maxwell's equations for the given geometry. The last step in this procedure is the application of boundary conditions at the interfaces, yielding the dispersion equations which must be solved to obtain the allowed modes. These dispersion equations are in general  {\em transcendental} and  except in a few simple cases, their solutions cannot be expressed in terms of elementary mathematical functions. 

When the solutions of the dispersion equations are known to be real, they are usually determined using a graphical search algorithm such as the bisection method \cite{numc}. However, in lossy material systems or low index-contrast asymmetric waveguides, the dispersion equations have complex solutions \cite{snyder_and_love}. This calls for a search in two dimensions (i.e., in the complex plane) and severely limits the effectiveness of graphical algorithms. 

Material loss and waveguide leakage are even more commonly encountered in plasmonic waveguides \cite{mark_book}. Due to the growing technological and scientific clout of plasmonic systems \cite{takahara:ol,weeber:prb,rashid:ol1,rashid:prb1,rashid:prb2,fan:apl1,fan:ol,bozhevolnyi:nat}various techniques for solving plasmon waveguide dispersion equations are in current use. Examples include the reflection-pole method (RPM) \cite{Anemogiannis1999,rashid:josaa}, Newton's method \cite{numc}, and the argument-principle method \cite{kocabas:prb}. Practical implementation of these methods requires careful programming customized for solving the problem at hand. In the methods that rely on curve-fitting, increasing the accuracy beyond a few decimals is often a challenging task, often requiring one to iterate the procedure manually  based on previous results. Additionally, the root-finding algorithms of several commercial mathematical software suites (e.g., \mtica) can be very sensitive to initial guesses provided by the user; estimation of this initial guess is especially difficult when the solution is complex.

In this paper, we present an easy-to-implement iterative procedure for solving complex transcendental dispersion equations that is relatively insensitive to initial guess. Our method applies to rectangular multilayer dielectric and plasmonic waveguides which may have either material or leakage loss. We first use a simple numerical example to illustrate the use of this technique and its convergence behavior. We then successively apply the procedure to find modes of dielectric slab waveguides, photonic wire waveguides, and plasmonic waveguides.

Papers in the past have discussed the use of iterative techniques for solving transcendental equations in general \cite{mckelvey:iterative}. Our aim in this paper is to demonstrate how this method can be applied to automate the design and analysis of waveguide structures of significant contemporary interest. 

\section{Iterative technique: A simple example}
Although the general philosophy of iterative methods is well known, we will use a simple example to introduce our terminology, illustrate the procedure use, and build an intuitive understanding of convergence issues. Suppose we need to solve the equation:
\begin{equation}\label{eq1a}
F(x)=0.
\end{equation}
where $F(x)$ is a combination of elementary mathematical functions. We reformulate this as an iterative problem by converting \myeqno{eq1a} to:
\begin{equation}\label{eq1b}
x=f(x).
\end{equation}
where $f$ is obtained by manipulating the parent function $F$. Next, we start with an initial guess $x_1$ and obtain a sequence according to:
\begin{equation}\label{eq1c}
x_{n+1}=f(x_n).
\end{equation}
If the sequence defined by \myeqno{eq1c} converges, then the limiting value is the solution to \myeqno{eq1b}. We will illustrate the method by a simple example. Consider the transcendental equation
\begin{equation}\label{ex1a}
\sin x=1-x.
\end{equation}
	\begin{figure}[!btp]
	\centering
	\includegraphics[width=5cm]{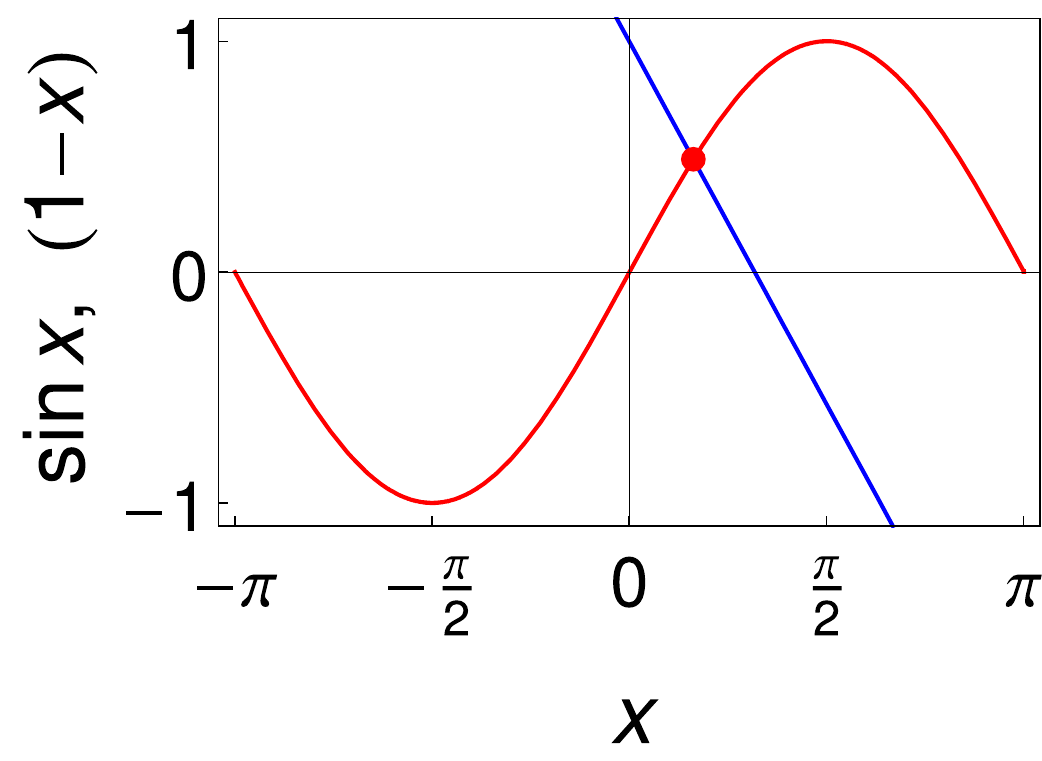}
	\caption{Graphs of the left- (red) and the right- (blue) hand sides of \myeqno{ex1a}. The red dot indicates the approximate location of the solution near $x\simeq0.5$.}
	\label{fig1}
\end{figure}
\myfigno{fig1} shows the graphs of the left-hand side (LHS) and right-hand side (RHS) of \myeqno{ex1a}, which intersect around $x\simeq0.5$. To find the root more accurately by the iterative method, we recast the equation in a form similar to \myeqno{eq1b} by choosing $f(x)=1-\sin x$, setting up an iteration scheme following \myeqno{eq1c}:
\begin{equation}\label{ex1b}
x_{n+1}=1-\sin x_n.
\end{equation}
\myfigno{fig2}(b) plots the first fifty iterations of \myeqno{ex1b} and indicates that the sequence converges to $x=0.5109734293885691$. We compare the 16-decimal agreement between this value and the solution $x=0.5109734293885691$ computed by \mtica's \texttt{FindRoot} function. \myfigno{fig2}(c) shows the LHS and RHS of the iteration scheme in \myeqno{ex1b} and \myfigno{fig2}(a) shows how the procedure converges to the intersection point of the two curves, starting from the initial guess. Making an initial guess corresponds to choosing a point on the curve $y=x$ shown by the red curve in \myfigno{fig2}(a). This point is indicated in \myfigno{fig2}(a) by the red dot. The vertical lines in the spiral represent computation of $f(x)$ for the chosen $x$  and the horizontal lines represent re-substitution of $x$ by $f(x)$ for the next iteration. 
	\begin{figure}[!tbp]
	\centering
	\includegraphics[width=13cm]{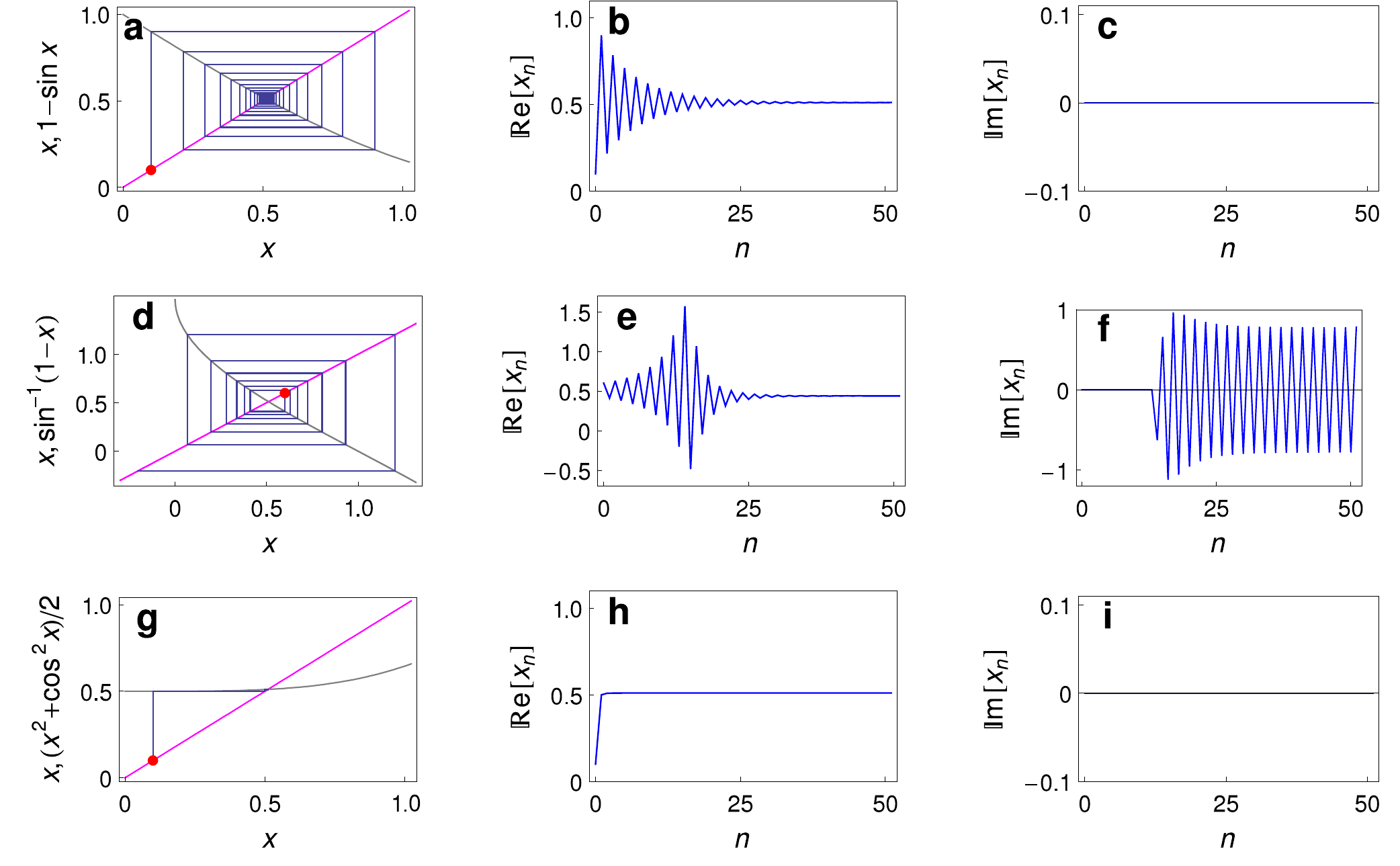}
	\caption{Solution of the example transcendental equation via the iterative method. The graphs (a), (d), and (g) show the left- (magenta) and the right- (gray) hand sides of \myeqno{ex1b}, \eqref{ex1c}, and \eqref{ex1d}. The red dot indicates the position of the initial guess. Convergence/divergence behavior of the real [(b), (e), (h)] and imaginary [(c), (f), (i)] parts of the iterates.}
	\label{fig2}
\end{figure}

Before we proceed to apply the iterative technique to practical waveguide problems, it is useful to gain an intuitive understanding of how the above scheme converges to the solution. This is important since, for any given $F(x)$, there are usually multiple ways to choose $f(x)$ which differ in their convergence behavior. For example, an equally-legitimate way of setting up an iteration scheme for \myeqno{ex1a} would be to choose $f(x)=\sin^{-1}(1-x)$ and obtain the sequence $\{x_n\}$ according to:
\begin{equation}\label{ex1c}
x_{n+1}=\sin^{-1}(1-x_n).
\end{equation}
 The behavior of successive iterates is shown in \myfigno{fig2}(d--f). Regardless of how close the initial guess is to the actual solution, the iterative scheme diverges (spirals away) from the intersection point. Although the real part of the solution appears to converge after about 25 iterations, the overall complex solution does not converge, as seen from the undamped oscillations in the imaginary part of the solution. Thus, this iteration scheme, although derived from the same parent equation, does not converge. This behavior is typical for transcendental equations involving trigonometric functions and is therefore encountered for waveguide problems, as we will show in section~\ref{wcdw}. 

On the other hand, convergence of \myeqno{ex1a} can be improved considerably by choosing an alternative iterative form as follows. Squaring both sides of \myeqno{ex1a} yields:
\begin{equation}\label{ex1d}
\sin^2 x=1-\cos^2x=(1-x)^2=1-2x+x^2.
\end{equation}
from which we can obtain yet another sequence of $\{x_n\}$ of iterations according to:
\begin{equation}\label{ex1e}
x_{n+1}=\frac{1}{2}\left(x_n^2+\cos^2x_n\right).
\end{equation}
\myfigno{fig2}(g--i) depict the convergence of succesive iterates of \myeqno{ex1e}. By comparing \myfigno{fig2}(b) with \myfigno{fig2}(h) it is apparent that \myeqno{ex1e} converges to the solution significantly faster than \myeqno{ex1b}.
	
\begin{figure}[!t]
	\centering
	\includegraphics[width=10cm]{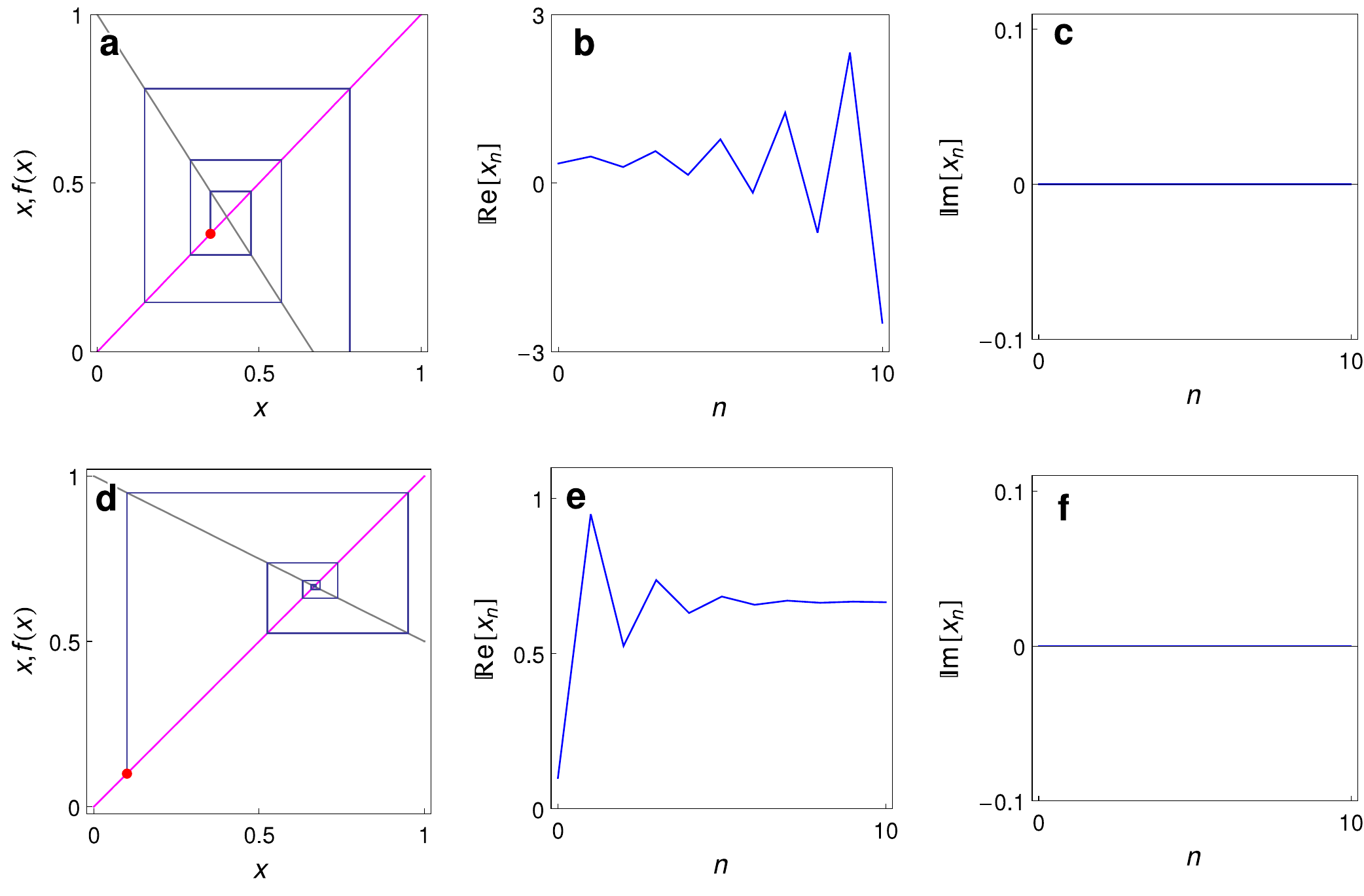}
	\caption{Criterion for convergence of the iterative solution. The magenta line in both figures is the curve $y=x$ and the blue lines are two different cases of $y=f(x)$. The solution (a) diverges  for $\absval{f'(x)}\geq1$ and (d) converges for $\absval{f'(x)}<1$. (b), (c), (e), and (f) show  how the convergence/divergence is reflected in the behavior of the real and the imaginary parts of the successive iterates.}
	\label{fig3}
	\end{figure}

 The preceding examples illustrate the crucial importance of choosing appropriate iterative forms. Arriving at such a form necessitates an understanding of the convergence of the iterative scheme. Figure~\ref{fig3} pictorially shows the criterion for convergence of the iterative scheme in \myeqno{eq1b}. Whether the iteration spirals toward the intersection point (the solution) or away from it is governed by the relative slopes of the intersecting curves. If $f(x)$ is steeper than $x$ (i.e., if $\absval{f'(x)}>1$), the successive computations and re-substitutions spiral away from the intersection and the solution diverges. On the other hand, if $f(x)$ rises gently compared to $x$ (i.e., if \mbox{$\absval{f'(x)}<1$}), then the scheme spirals toward the intersection and the solution converges. For convergent functions, the rate of convergence is governed by the magnitude of $\absval{f'(x)}$. The RHS of \myeqno{ex1e} is much flatter ($ \absval{f'(x)}\ll1$) near the solution compared to the RHS of \myeqno{ex1b}, which results in the latter's considerably slower convergence. While the preceding justification is not a rigorous proof, it provides a basis for choosing the manipulations required to obtain convergent forms of the practically-useful equations we consider in the forthcoming sections \cite{fptheory}. 
\section{Dispersion equation of a general asymmetric three-layer structure}\label{dpmodes}
\begin{figure}[tb]
	\centering
	\includegraphics[width=12cm]{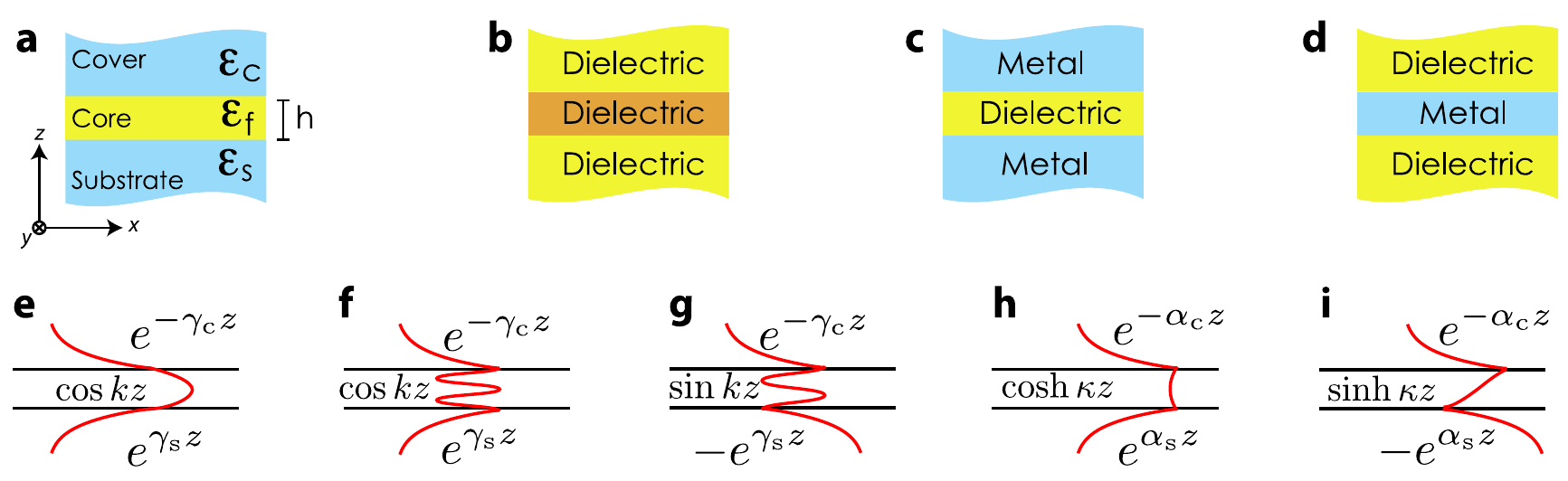}
	\caption{Geometries and modes of three-layer infinite slab waveguide structures. Parts (e--i) plot the typical magnetic field profiles.}
	\label{fig4}
	\end{figure}

 The design of many integrated photonics systems of practical importance---including both novel plasmonic waveguide architectures as well as conventional dielectric waveguides---can often be reduced to solving for the effective mode index and electromagnetic field distributions in a three-layer planar structure.  As such, we focus our analysis on a generalized three-layer slab waveguide, as shown in \myfigno{fig4}(a).  The central layer is called the core and has a complex relative permittivity $\epsf$. The bottom and top cladding layers are called the substrate (with permittivity $\epss$) and cover (with permittivity $\epsc)$, respectively. 

Depending on which materials comprise the three layers, it is possible to classify the most  commonly-encountered waveguides into three basic types. The dielectric waveguides, shown in \myfigno{fig4}(b), have all three layers made of dielectrics, with $\epsf>\epss,\epsc$. The structures in Figs.~\ref{fig4}(c) and (d) are two basic configurations of plasmonic waveguides. The first, shown in \myfigno{fig4}(c), consists of an dielectric sandwiched between two (possibly different) metals and is commonly known as a metal-dielectric-metal (MDM) waveguide. The second structure, shown in \myfigno{fig4}(d), consists of a metal layer between two (possibly different) dielectrics; it is commonly known as an dielectric-metal-dielectric (DMD) waveguide. 

Figures~\ref{fig4}(e--i) schematically show the possible modes of a general three-layer slab waveguide. These modes are distinguished according to several criteria: mode number (number of zero-crossings in the field), polarization (transverse electric or transverse magnetic), field symmetry (even or odd), and field profile in the core (sinusoidal or hyperbolic). Of the possible field profiles shown in \myfigno{fig4}, the mode shown in (e), having sinusoidal core-field variation with no zero-crossings, is exclusive to dielectric waveguides and is known as the fundamental mode. Modes (h) and (i), having hyperbolic core-field variation, are exclusive to plasmonic waveguides and are known as the gap-plasmon modes. Modes (f) and (g), having sinusoidal core-field variation with $\geq1$ zero-crossings, are common to both dielectric and plasmonic waveguides.  

This rich variety of modes arises out of the solutions of the dispersion equation written for an asymmetric three-layer slab waveguide. In principle, a single dispersion equation can completely describe all the modes of both dielectric and plasmonic waveguides. However, because the nominally-assumed core-field distributions for the two types are different (sinusoidal for dielectric waveguides and hyperbolic for plasmonic waveguides), we choose to write two separate equations for the two types \cite{pollock:book,burke:prb}: 
\begin{align}
\label{dwdisp}
\tan(kh)&=\frac{k(p\gc+q\gs)}{k^2-pq\gc\gs}&\ldots\text{for dielectric waveguides.}\\ \nonumber \\
\label{pwdisp}
\tanh(\kappa h)&=-\frac{\kappa(p\ac+q\as)}{\kappa^2+pq\ac\as}&\ldots\text{for plasmonic waveguides.}
\end{align}
For convenience, we have collected the symbol definitions and their expressions in \mytabno{tab1}. Using these definitions, Eqs.\ \eqref{dwdisp} and \eqref{pwdisp} can be cast explicitly as transcendental equations in a single complex variable $k$ or $\kappa$. Furthermore, it is evident that \myeqno{dwdisp} transforms to \myeqno{pwdisp} for purely imaginary values of $k$ (i.e., $k\to i\kappa,\,\kappa\in\mathbb{R}$).  Having solved for $k$ or $\kappa$, the mode's complex effective index $\neff$ may be calculated using relations in \mytabno{tab1}.
\begin{table}[!tb]
\centering
	\begin{tabular}{c|c}
	\hline
	{\bf Symbol}&{\bf Definition/Expression}\\  \hline \hline
	$k_0$&$2\pi/\lambda_0$\\ \hline
	 $p$&1 (for TE), $\epsf/\epsc$ (for TM)\\ \hline
	$q$&1 (for TE), $\epsf/\epss$ (for TM)\\ \hline
	$\Kc$, $\Ks$&$k_0\sqrt{\epsf-\epsc}$, $k_0\sqrt{\epsf-\epss}$\\ \hline
	$\Qc$, $\Qs$&$k_0\sqrt{\epsc-\epsf}$, $k_0\sqrt{\epss-\epsf}$\\ \hline
	$\gc$, $\gs$&$\sqrt{\Kc^2-k^2}$, $\sqrt{\Ks^2-k^2}$\\ \hline
	$\ac$, $\as$&$\sqrt{\Kc^2+\kappa^2}$, $\sqrt{\Ks^2+\kappa^2}$\\ \hline
	$\xc$, $\xs$&$\sqrt{\kappa^2-\Qc^2}$, $\sqrt{\kappa^2-\Qs^2}$\\ \hline
	$\Gc$, $\Gs$&$\sqrt{k^2+p^2\gc^2}$, $\sqrt{k^2+q^2\gs^2}$\\ \hline
	$S$&$(p\ac+q\as)/2$\\ \hline	
	$A$, $B$&$(p\xc+q\xs)/2$, $(p\xc-q\xs)/2$\\ \hline
	 $\neff$&$\sqrt{\epsf-\left(k/k_0\right)^2}$ (sinusoidal core-fields),\\
	 &$\sqrt{\epsf+\left(\kappa/k_0\right)^2}$ (hyperbolic core-fields)\\ \hline		
	\end{tabular}
	\caption{Definitions of various quantities and their expressions in terms of the material parameters and the perpendicular core wavevector $k$ or $\kappa$. }
\label{tab1}
\end{table}
	\begin{figure}[tb]
	\centering
	\includegraphics[width=5cm]{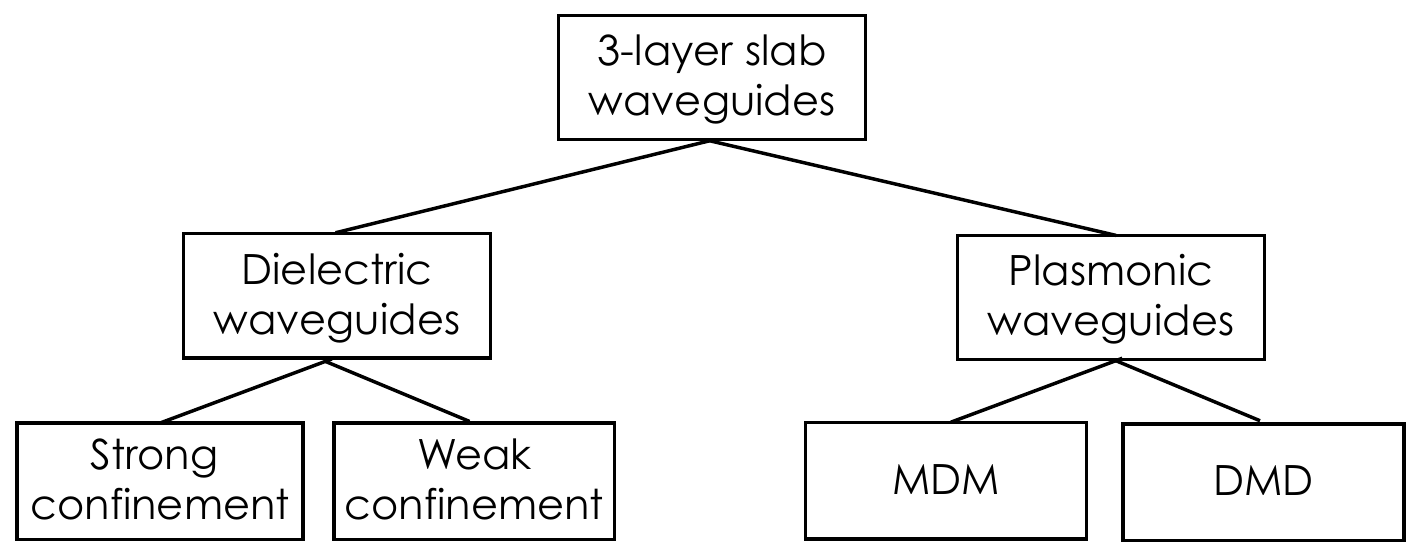}
	\caption{Categories of three-layer slab waveguides for the purposes of iterative solution.}
	\label{fig5}
	\end{figure}

We will now begin the description of the actual iterative solution procedure. We saw in section \ref{intro} that the convergence of this technique depends crucially on the iteration function. As a result, the iterative form for determining modes with sinusoidal core-fields [\myfigno{fig4}(e--g)] is different from the one needed for determining modes with hyperbolic core-fields [\myfigno{fig4}(h--i)]. Therefore, we split our description into two broad categories: dielectric waveguides and plasmonic waveguides. These two categories are further divided depending upon the degree of confinement (strong or weak) for dielectric waveguides, and core/cladding type (MDM or DMD) for plasmonic waveguides. Figure \ref{fig5} depicts the division of the waveguide modes that we have made for the purposes of describing our iterative solution process. In the forthcoming sections, we will obtain and test rapidly-convergent iterative forms for calculating the mode indices of these four sub-categories of three-layer asymmetric slab waveguides. These useful forms will be boxed for the convenience of the reader. 
\section{Modes of dielectric waveguides}
\subsection{ Strong-confinement dielectric waveguides}
Using the half-angle identity $\tan(z/2)=-\cot z\pm\sqrt{1+\cot^2z}$, \myeqno{dwdisp} can be transformed to:
\begin{equation}\label{sw2}
\tan(kh/2)=\frac{\left(pq\gc\gs-k^2\right)\pm\Gc\Gs}{k(p\gc+q\gs)}.
\end{equation}
Inverting the tangent function in \myeqno{sw2} gives us the convergent iterative form for strong-confinement waveguides as:
\begin{equation}\label{sw2it}
\boxed{k_{n+1}=\frac{2}{h}\left\{M\pi+\tan^{-1}\left[\frac{\left(pq\gcn\gsn-k_n^2\right)\pm \Gcn\Gsn}{k_n(p\gcn+q\gsn)}\right]\right\}}
\end{equation}
Here $M$ is an integer and the quantities $\gcn$, $\gsn$, $\Gcn$, and $\Gsn$ are related to the $n^{\mathrm{th}}$ iterate $k_n$ through the relationships in \mytabno{tab1}.  Although the form of \myeqno{sw2it} may seem daunting, we note that it is a general equation which can calculate odd and even modes of both transverse electric (TE) and transverse magnetic (TM) polarizations for asymmetric slab waveguides. All of the functions involved in \myeqno{sw2it} are found on scientific calculators and in mathematical software tools.  The iterative procedure obviates any need for a graphical search and has an inherent ability to give arbitrary-precision solutions---a feat difficult to achieve using graphical interpolation. 

Before embarking on actual computation, we need to specify the parity ($+$ or $-$ sign), order ($M$), and the polarization ($p$ and $q$) of the desired mode. For TE polarization ($p=q=1$) and even parity ($+$ sign), the effective indices of successive even TE modes may be simply calculated by setting $M=0, 1, \dots m$ and iterating according to \myeqno{sw2it}; this yields the TE$_0$, TE$_2$, $\dots$ TE$_{2m}$ modes. For TE polarization with odd parity ($-$ sign), setting $M=1, 2, \dots, m$ yields the TE$_1$, TE$_3$, $\dots$ TE$_{2m-1}$ modes. Note that for even modes, the mode order begins with $M=0$,  whereas for odd modes, it begins with $M=1$. The procedure for obtaining TM mode indices is identical, with proper input of $(p,q)$ as shown in Table \ref{tab1}.
 
\begin{figure}[!tb]
	\centering
	\includegraphics[width=13cm]{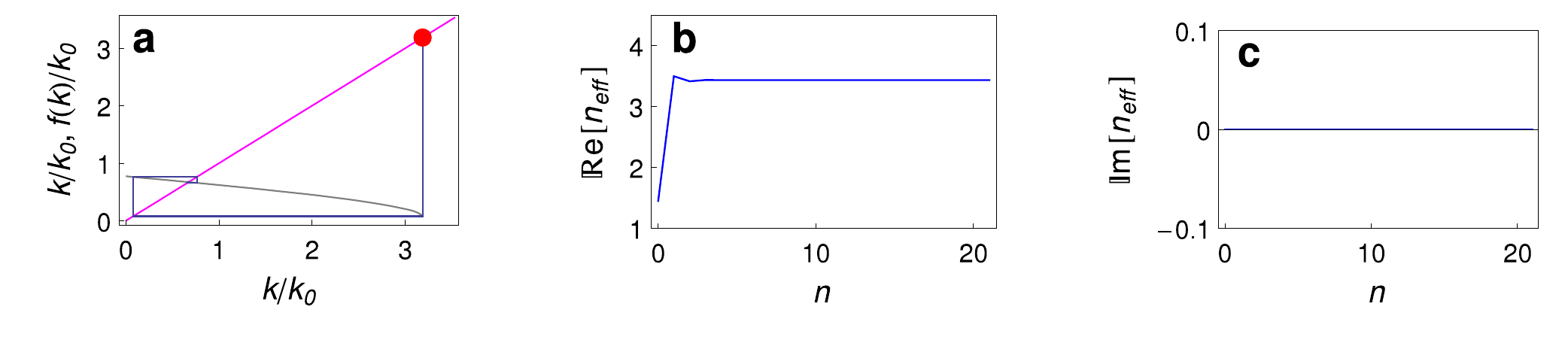}
	\caption{Iterative solution of a strong-confinement waveguide using \myeqno{sw2it}. (a) The convergence plot depicting the normalized LHS (magenta) and RHS (gray) of \myeqno{sw2it}; $f(k)$ on the $y$-axis of of (a) refers to the function on the right hand side. The convergence of the real and imaginary parts of the effective index are shown in (b) and (c), respectively.}
	\label{fig6}
	\end{figure}

To illustrate the use of the iterative method, we choose a 1-$\mum$-thick silicon-on-insulator (SOI) slab waveguide operating at 1550 nm as a test structure. The refractive indices of the silicon film, oxide substrate, and air cover are assumed to be $\sqrt{\epsf}=3.50, \sqrt{\epss}=1.45$, and $\sqrt{\epsc}=1.00$ respectively.   Using the TE$_0$ mode as an example, \myfigno{fig6} depicts how \myeqno{sw2it} converges to the solution. \myfigno{fig6}(b) and (c) show the real and imaginary values of the first twenty iterates of \myeqno{sw2it}. Although the iteration is carried out in terms of $k$, we have chosen to depict the convergence in terms of the effective index $n_{\mathrm{eff}}=\beta/k_0$ since this is a more familiar quantity in waveguide analysis. \myfigno{fig6}(a) shows the LHS and RHS of \myeqno{sw2it} and how the iterative scheme converges to the solution. Notice that the slope condition mentioned in section \ref{intro} is satisfied for this particular case. 
	
To examine the iterative technique in contrast with standard numerical techniques, we solved for the modes of the same structure using \mtica{}.\mytabno{tab2} compares the two solutions. \mtica{} calculations were performed by feeding \myeqno{sw2} to the \texttt{FindRoot} function with an initial guess dependent on the mode order, parity, and polarization. To obtain the initial guesses, we plotted the left- and right-hand sides of \myeqno{sw2} and made a manual estimate based on  the intersection of the graphs. Given the multiple solutions inherent in \myeqno{sw2}, these initial guesses needed to be fairly close to the actual solution for \mtica{} to return the correct solution. On the other hand, all iterative solutions in \mytabno{tab1} were initiated with the same initial guess $k_1=\Ks$. Moreover, the convergence of the iterative technique is completely insensitive to the exact value of this initial guess: we obtained the same effective indices for initial guesses ranging from  $\Ks$ to $10^{5}\Ks$ (including complex values). While a precise mathematical characterization of the convergence behavior is outside the scope of our paper, this exercise does suggest the robustness of the iterative technique with respect to the accuracy of the initial guess.  
	\begin{table}[!bt]
	\centering
		\begin{tabular}{c|c|c|c|c|c}
		\hline 
		{\bf Mode}&$(p,q)$&{\bf Parity}&$M$&{\bf Iterative scheme}&{\bf \mtica{}}\\  \hline \hline
		TE$_0$&(1,1)&+&0&3.4347458991523551&3.4347458991523551\\ \hline
		TE$_1$&(1,1)&-&1&3.2327892969869200&3.2327892969869201\\ \hline
		TE$_2$&(1,1)&+&1&2.872310278807719&2.872310278807719\\ \hline
		TE$_3$&(1,1)&-&2&2.302024617480549&2.302024617480549\\ \hline
		TM$_0$&$(\epsf/\epsc,\epsf/\epss)$&+&0&3.4165068626393461&3.4165068626393461\\ \hline
		TM$_1$&$(\epsf/\epsc,\epsf/\epss)$&-&1&3.1541909024008027&3.1541909024008027\\ \hline
	       TM$_2$&$(\epsf/\epsc,\epsf/\epss)$&+&1&2.668932488161409&2.668932488161409\\ \hline
		TM$_3$&$(\epsf/\epsc,\epsf/\epss)$&-&2&1.865243634178012&1.865243634178012\\ \hline
		\end{tabular}
	\caption{Comparison of mode indices for a silicon-on-insulator (SOI) slab waveguide computed via the iterative method and the \texttt{FindRoot} function in \mtica{}}
	\label{tab2}
	\end{table}
	
\subsubsection{Symmetric strong-confinement waveguides}
If the relative permittivities of the cover and the substrate are equal, then the structure is called a symmetric waveguide. Since several important photonic device structures employ symmetric waveguides, we now give explicit iterative forms for calculating their mode indices. These iterative forms arise out of the simplifications in \myeqno{sw2it} due to the equality of $\epss$ and $\epsc$: 
\begin{subequations}
\begin{align}\label{scit}
k_{n+1}&=\frac{2}{h}\left[M\pi+\tan^{-1}\left(p\sqrt{\Kc^2/k_n^2-1}\right)\right]\quad\text{\ldots for even modes.}\\
k_{n+1}&=\frac{2}{h}\left[M\pi-\cot^{-1}\left(p\sqrt{\Kc^2/k_n^2-1}\right)\right]\quad\text{\ldots for odd modes.}
\end{align}
\end{subequations}
We also note that, like \myeqno{sw2it}, $M$ assumes values starting from $0$ for even modes and $1$ for odd modes. 

\subsection{Weak-confinement dielectric waveguides}\label{wcdw}
\myeqno{sw2it} converges for a very wide range of refractive indices, wavelengths, and waveguide thicknesses which are commonly used in practical integrated photonics designs. It performs poorly, however, when applied to the weak-confinement single-mode waveguide structures routinely fabricated out of III-V (e.g., GaAs/AlGaAs) or organic materials. In the context of multimode waveguides, the strong-confinement formula encounters convergence problems when used to calculate indices of modes near the waveguide cutoff. We highlight this limitation and its remedy through the following example. 

\begin{figure}[!tb]
	\centering
	\includegraphics[width=13cm]{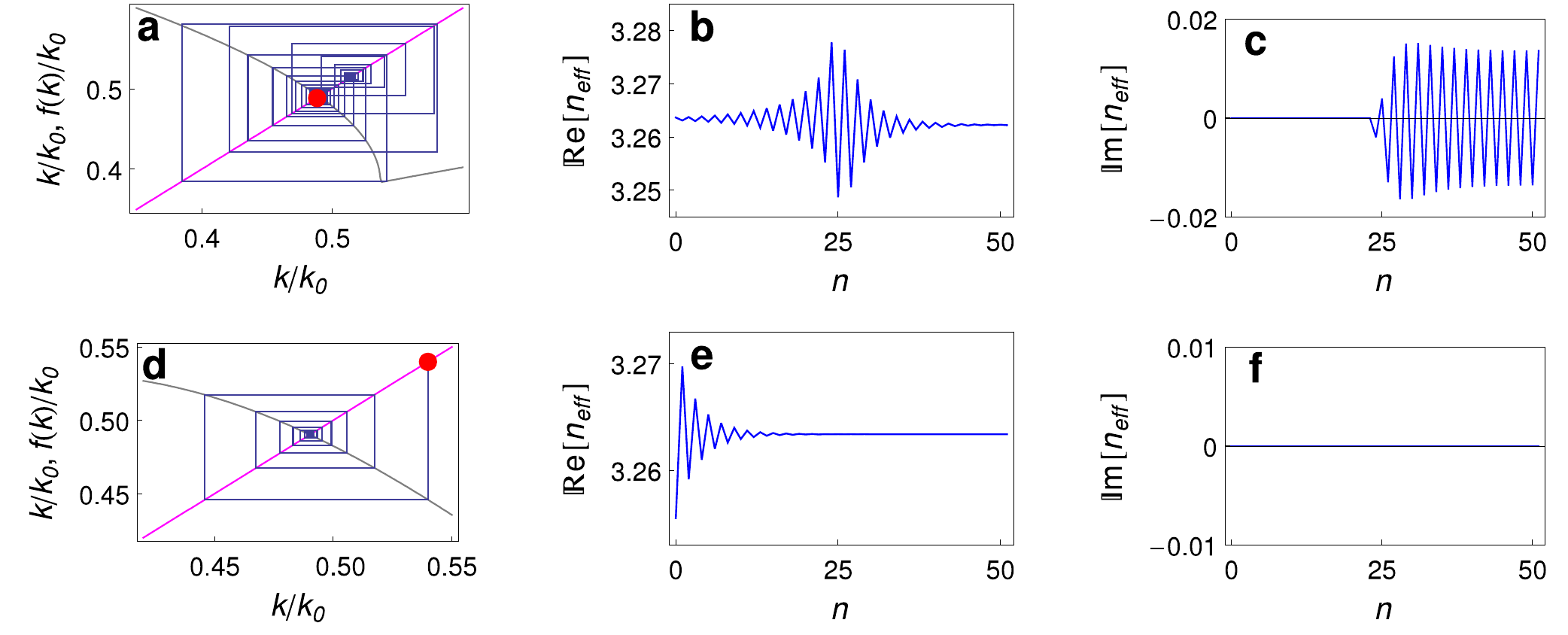}
	\caption{(a--c) Iterative solution of a weak-confinement waveguide using \myeqno{sw2it}. (a) The LHS and RHS of \myeqno{sw2it}; $f(k)$ on the y-axis denotes the RHS of \myeqno{sw2it}. Notice the apparent convergence of the real part (b) and the oscillatory divergence in the imaginary part (c) of the effective index. (d--f) Solution of the same weak-confinement waveguide problem using \myeqno{sw3it}. (d) The LHS and RHS of \myeqno{sw3it};  $f(k)$  on the y-axis denotes the RHS of \myeqno{sw3it}. (e) and (f) show the convergence of the real and imaginary parts of the effective index.}
	\label{fig7}
	\end{figure}

Consider a 1-$\mum$-thick air-clad GaAs waveguide on an Al$_{0.1}$Ga$_{0.9}$As substrate operating at 1550 nm. The refractive indices of GaAs and Al$_{0.1}$Ga$_{0.9}$As are assumed to be $\sqrt{\epsf}=3.300$ and $\sqrt{\epss}=3.256$ respectively. We attempt to find the fundamental TE mode iteratively by inputting the corresponding parameters to \myeqno{sw2it} ($M=0$, $p=q=1$, and $+$ sign). \myfigno{fig7}(b) and (c) show the behavior of the real and imaginary parts of the calculated effective index for first 50 iterates. The real part of the effective index initially diverges to reach a maximum at iteration number 30 and begins to converge thereafter. Starting from the thirtieth iteration,  the imaginary part oscillates between $\pm0.01$. These oscillations do not damp out even if we increase the number of iterations to $10^4$. \mtica{}, however, indicates that this structure supports TE$_0$ and TM$_0$ modes with indices of 3.266 and 3.263, respectively. We conclude that the iterative scheme in \myeqno{sw2it} fails to converge for this example. 

This failure to converge can be understood by examining the behavior of the LHS and RHS of \myeqno{sw2it} near the intersection point, as shown in \myfigno{fig7}(a). Near the solution, $\absval{f_1'(\kappa)}>1$. As a result, the iterates begin to diverge away from the intersection point. In fact, $\absval{f_1'(\kappa)}$ continues to increase away from the intersection, causing a rapidly-increasing divergence. However, at $\kappa\simeq0.537k_0$, $\absval{f_1'(\kappa)}$ abruptly becomes $<1$, causing the real part of the iteration to converge. This convergence of the real part is misleading, however since the imaginary part shows undamped oscillations. The behavior of the strong confinement formula in this case is similar to the example problem in section \ref{intro} illustrated in \myfigno{fig2}(d--f).

In summary, the basic cause of the failure of convergence is the breakdown of the slope condition mentioned in section \ref{intro}. This is remedied by recasting \myeqno{dwdisp} into a form which satisfies the convergence condition. To this end, we use the identity $\tan^2 z=1/\cos^2z-1$, rewriting \myeqno{dwdisp} as: 
\begin{equation}\label{sw3}
k^2\pm\Gc\Gs\cos(kh)=pq\gc\gs.
\end{equation}
	\begin{table}[!bt]
	\centering
		\begin{tabular}{c|c|c|c}
		\hline 
		{\bf Mode}&$(p,q)$&{\bf Iterative scheme}&{\bf \mtica{}}\\  \hline \hline
		TE$_0$&(1,1)&3.26599646645606654&3.26599646645606654\\ \hline
		TM$_0$&$(\epsf/\epsc,\epsf/\epss)$&3.26338400537407312&3.26338400537407312\\ \hline
		\end{tabular}
	\caption{Comparison of mode indices for an Al$_{0.1}$Ga$_{0.9}$As/GaAs/air slab waveguide computed via the iterative method and the \texttt{FindRoot} function in \mtica{}}
	\label{tab3}
	\end{table}
The convergent form suitable for weak confinement is obtained by solving for $k$ contained in the $\gc\gs$ term on the right hand side of \myeqno{sw3}:
\begin{equation}\label{sw3it}
\boxed{k_{n+1}=\sqrt{\frac{(pq\Kc\Ks)^2-(\Gcn\Gsn)^2\cos^2(k_nh)+(p^2q^2-1)k_n^4}{p^2q^2\left(\Kc^2+\Ks^2\right)\mp2\Gcn\Gsn\cos(k_n h)}}}
\end{equation}
where $\Gcn$ and $\Gsn$ are related to $k_n$ through the relationships in \mytabno{tab1}. We emphasize that \myeqno{sw3it} is a general equation capable of solving for near-cutoff TE and TM modes of odd and even parity in three-layer asymmetric slab waveguides. The choice of $+$ or $-$ sign in \myeqno{sw3it} depends on whether the cutoff occurs at the odd $(+)$ or the even $(-)$ mode. When analyzing a given structure, it is difficult to determine {\em a priori} the parity of the cutoff mode. The prescription is therefore to use the strong-confinement formula until it fails to converge, continuing with the weak-confinement formula thereafter. The domain of convergence of the strong- and weak-confinement formulas may be determined rigorously by setting the derivatives of the RHS of Eqs.~\eqref{sw2it} and \eqref{sw3it} equal to 1 and solving for the corresponding value of $k$.

Solving for the mode indices of the previously-considered GaAs/Al$_{0.1}$Ga$_{0.9}$As waveguide using \myeqno{sw3it} yields the effective indices of  the TE$_0$ and TM$_0$ modes in excellent agreement with the \mtica{} results, as seen from \mytabno{tab3}. The convergence behavior of the effective index is plotted in \myfigno{fig7}(d--f). The convergence of \myeqno{sw3it} improves as the core--cladding index contrast decreases; similarly, the convergence of \myeqno{sw2it} improves for {\em increasing} core--cladding index contrast. 
\subsubsection{Symmetric weak-confinement dielectric waveguides}
Similarly to strong-confinement waveguides, having identical cover and substrate indices considerably simplifies the iterative form for weakly-confined slab waveguides, which can be written as:
\begin{subequations}
\begin{align}\label{wcit}
k_{n+1}&=\frac{\Kc}{\sqrt{1+p^{-2}\tan^2\left(k_nh/2\right)}}&\quad\text{\ldots for even modes.}\\
k_{n+1}&=\frac{\Kc}{\sqrt{1+p^{-2}\cot^2\left(k_nh/2\right)}}&\quad\text{\ldots for odd modes.}
\end{align}
\end{subequations}

\subsection{Extension to photonic wire waveguides}
 Waveguides  in which light is confined in two dimensions and propagates in the third dimension are known as photonic wire (PW) waveguides;  optical fibers, as well as ridge, rib, and channel waveguides are examples. Because they offer stronger light confinement compared to slab waveguides, PW waveguides are commonly employed in on-chip optical devices where a small size is essential for achieving several critical device specifications ( e.g., speed, low power, integration density, etc.). 
	\begin{figure}[!tb]
	\centering
	\includegraphics[width=4in]{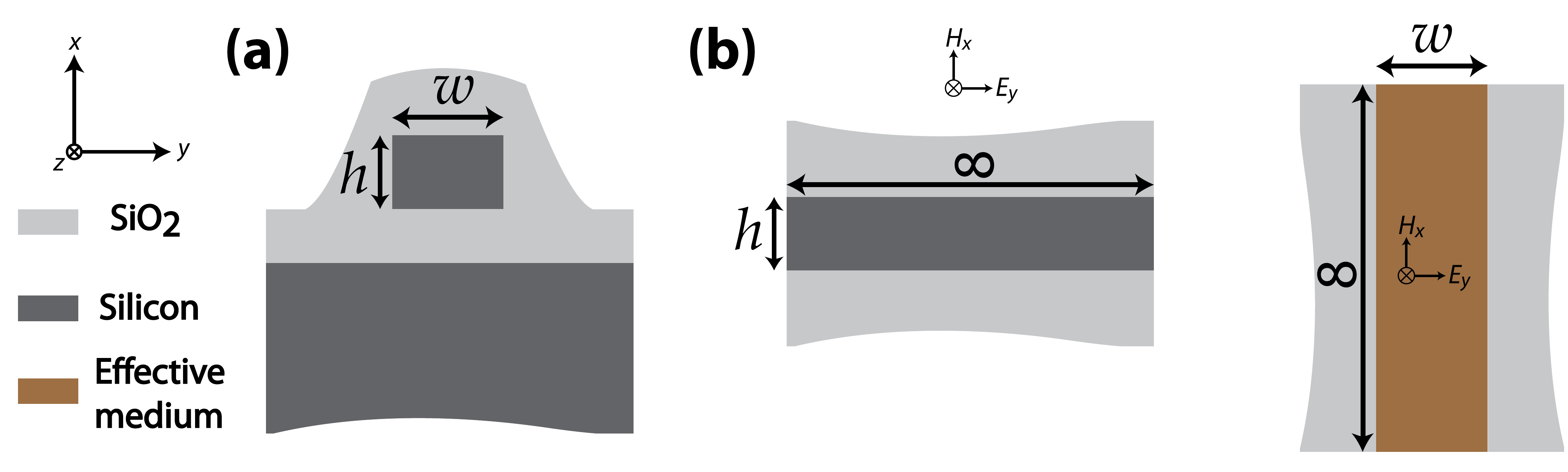}
	\caption{(a) Typical geometry of a SOI photonic wire waveguide. (b) The two steps used in determining the mode index of the PW waveguide using the effective index method. The procedure is illustrated here for a $E_y$-polarized mode; steps are similar for a $H_y$-polarized mode.}
	\label{fig8}
	\end{figure}

Although the exact field distribution in PW waveguides is best determined by numerical calculations, the effective index method (EIM) is remarkably successful at quickly calculating the mode indices for several common PW waveguide configurations \cite{pollock:book}. A knowledge of the effective index suffices for many important design problems. Figure~\ref{fig8} illustrates the two steps in the implementation of EIM. Figure~\ref{fig8}(a) shows an idealized silica-clad SOI photonic wire waveguide having a width $w$ and thickness $h$. In the first step, we disregard the confinement in the $x$-direction and solve for the effective index $n'$. Depending on the polarization of the desired mode, we will need to solve either the TE or TM equation. In the second step, we consider a silica-clad slab waveguide with a core index $n'$ confined in the $x$-direction. Note that the orientation of the effective waveguide is orthogonal to that in the first step and hence the equation in this step is for the polarization orthogonal to that in the previous step. That is, if the first step uses the TE dispersion equation, then the second step uses the TM equation(and vice versa). The effective index $n''$ determined in this second step is our final answer for the effective mode index of the original PW waveguide. 
	\begin{figure}[!b]
	\centering
	\includegraphics[width=4in]{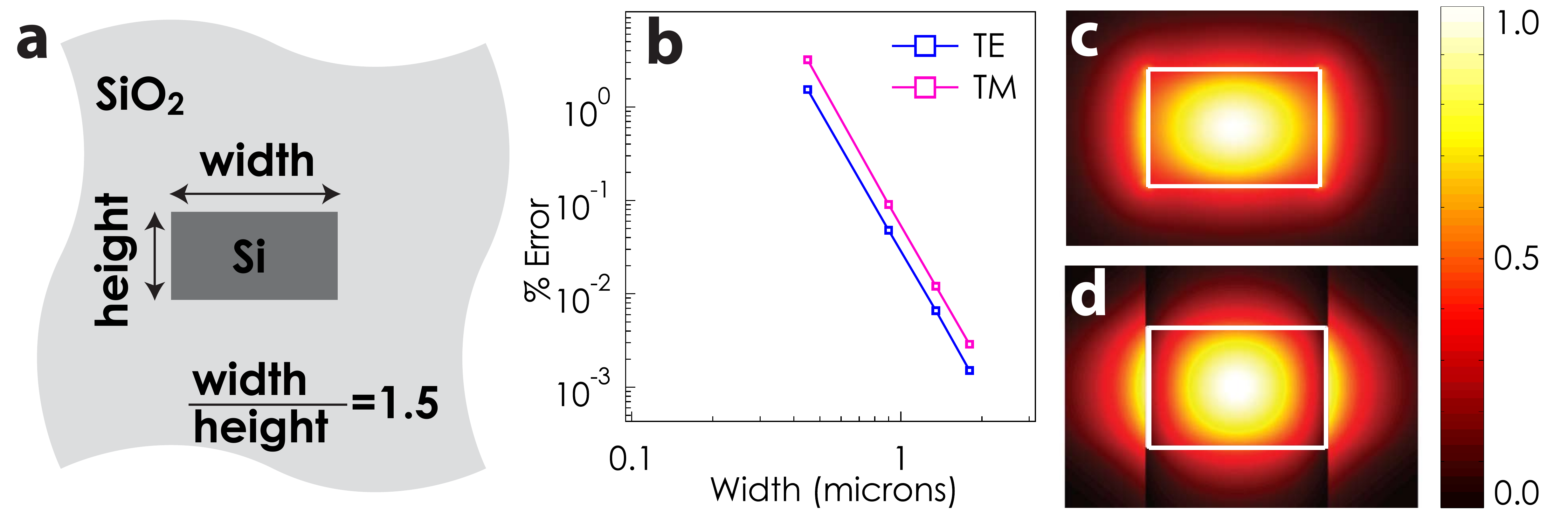}
	\caption{(a) Photonic wire waveguide structure used for evaluation of the effective index method (EIM). (b) The relative error in the calculation of the mode index using EIM as a function of the waveguide width. Relative error is defined as $\absval{n''\ssub{iter}-n''\ssub{FEM}}/n''\ssub{FEM}$. The electric field of the fundamental mode of a SOI photonic wire waveguide calculated using (c) the finite element method and (d) the EIM.}
	\label{fig9}
	\end{figure}

As an example, suppose that we need to determine the effective index of the fundamental $E_y$-polarized mode of a sub-micron PW waveguide shown in \myfigno{fig9} (a) with $h=450\mnm$ and $w=300\mnm$. In the first step, we solve for the TE mode equation for the structure shown in \myfigno{fig8}(b). Since SOI waveguide constitutes a strong-confinement system, the first step is readily accomplished by iteration of \myeqno{sw2it} for the TE condition (namely, $M=0$, $p=q=1$). This yields the first effective index as $n'=3.073930677459340$. In the second step, we consider an $x$-confined slab with core-index $n'$ and iterate \myeqno{sw2it} with the fundamental TM mode condition. This yields the effective index of the PW waveguide as $n''=2.652766507502340$. 

To test the relative accuracy of the iterative method, we use the finite element method (FEM) to compute the mode index. For the waveguide with the above dimensions, FEM gives $n''=2.612594$. It is evident that even for sub-mircron dimensions, the relative error defined as $\absval{n''\ssub{iter}-n''\ssub{FEM}}/n''\ssub{FEM}$ is only 1.5\%. Figure~\ref{fig9}(b) shows the variation of the relative error with waveguide dimensions for a fixed width-to-height aspect ratio of 1.5, from which we note the exponential drop in the relative error with increasing waveguide size. The reason for the decrease of accuracy with decreasing size lies in the increased interaction of the electromagnetic fields with the corner regions of the waveguide. By decomposing a two-dimensional mode-solving problem into two one-dimensional problems, we effectively chose to neglect the interaction of the fields with the  waveguide corners. This assumption becomes decreasingly accurate with decreasing waveguide size. Figure~\ref{fig9}(c) and (d) show the $y$-component of the electric field, calculated using FEM and the iterative method, for the sub-micron waveguide considered above. Note the strong electric field at the waveguide corners in \myfigno{fig9}(c) calculated using FEM, and its absence in \myfigno{fig9}(d) calculated using the iterative method. In addition to highlighting this difference for small waveguides, this exercise illustrates the well-known fact that while EIM successfully calculates the effective indices, it does not guarantee satisfaction of the boundary conditions. It should therefore not be used to calculate fields for wavelength-sized structures.

In spite of this limitation, the EIM remains an intuitive way of accomplishing designs which rely primarily on the knowledge of mode indices. Examples of such designs include several contemporary device structures such as SOI ring resonators, Bragg gratings, Mach-Zehnder interferometers, and directional couplers. By offering a simple means of solving the slab waveguide dispersion equation, the iterative method greatly enhances the efficacy and implementation speed of EIM. 
\section{Modes of plasmonic waveguides}
The examples in the preceding sections indicate the relative simplicity of the iterative method when applied to dielectric waveguides. However, the true advantage of this technique emerges when dealing with systems having material loss (complex permittivity) or leakage loss (complex propagation constant). The modal indices are complex in both these cases and providing graphical root-finders with a complex initial guess becomes difficult. As such, the self-converging behavior of the iterative algorithm becomes an invaluable asset. 

Propagation loss and waveguide leakage arise routinely in sub-wavelength-scale metallic waveguides. Intense research efforts are currently underway in the field of metal-based optics, also known as plasmonics (see, e.g., \cite{mark_book} and references contained therein). Extended metal structures support surface plasmon-polariton (SPP) modes that are electromagnetic waves strongly coupled to collective electron oscillations in the metal. To exploit  the strong light localization achievable in plasmonic structures, a variety of waveguide configurations have recently been proposed and demonstrated \cite{mark_book}. However, because of its simplicity, the MDM geometry remains a canonical structure for achieving and studying strong light confinement. As such, a convenient technique for determining the optical modes of MDM waveguides is highly valuable. 

We begin the description of the iterative solution technique for plasmonic waveguides by referring to the division of slab waveguide types illustrated in \myfigno{fig4}(c) and (d). We first consider the modes of an MDM-type plasmonic waveguide followed by a treatment of the DMD-type. 

\subsection{Metal-dielectric-metal waveguides}
A MDM waveguide supports gap-plasmon  and TM-like waveguide modes. Even and odd gap-plasmon modes are practically important, as they offer the strongest sub-wavelength field confinement. However, TM-like modes are also of theoretical value and are necessary for gaining a complete understanding of reflection and transmission phenomena in MDM waveguides and antennas \cite{kocabas:prb}. Effective indices for both these types of modes can be conveniently determined using the iterative technique, as we show in the following.

\subsubsection{ Gap-plasmon modes}
We start with our master plasmonic \myeqno{pwdisp} and rearrange it as:
\begin{equation}\label{pw1}
\kappa^2+2S\kappa\coth(\kappa h)+pq\ac\as=0.
\end{equation}
Here, $\ac$,$\as$, and $S$ are related to $\kappa$ as specified in \mytabno{tab1}. Treating this as a quadratic equation in $\kappa$ and solving yields the convergent iterative form for the MDM gap-plasmon modes as:
\begin{equation}\label{pw1it}
\boxed{\kappa_{n+1}=-S_n\coth(\kappa_n h)\pm\sqrt{S_n^2\coth^2(\kappa_n h)-pq\acn\asn}}
\end{equation}
	\begin{figure}[b]
	\centering
	\includegraphics[width=13cm]{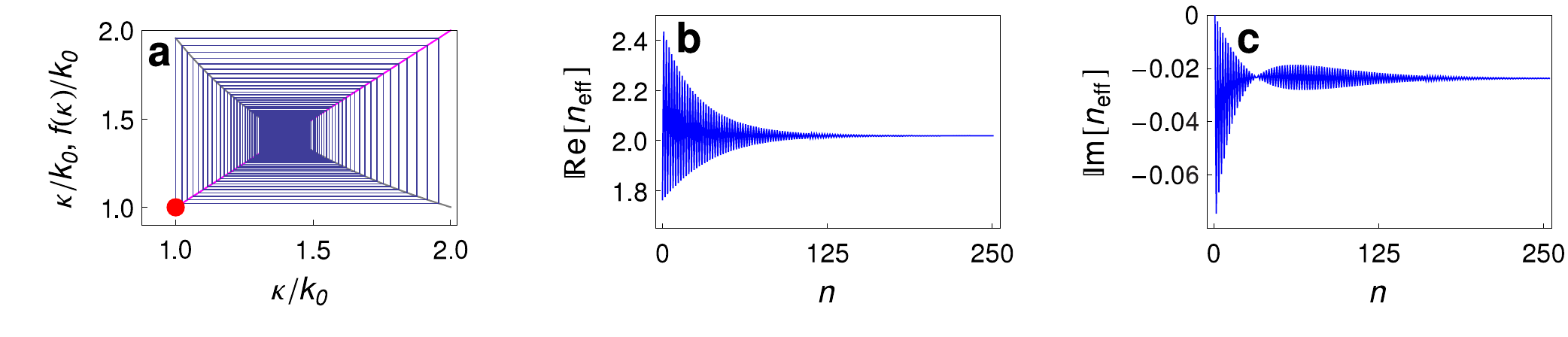}
	\caption{(a) Normalized left- and right-hand sides of \myeqno{pw1it}; $f(\kappa)$ refers to the RHS. Convergence of the real (b) and the imaginary (c) parts of the effective index for the fundamental gap-plasmon mode.}
	\label{fig10}
	\end{figure}
This simple-looking equation can calculate even ($+$ sign) and odd ($-$ sign) gap-plasmon mode indices of a wide variety of deep sub-wavelength asymmetric plasmonic waveguides. We illustrate the use of \myeqno{pw1it}  through two examples. Our first structure is a 50-nm-thick gold-silica-silver slab waveguide operating at 1550 nm. The relative permittivities of gold, silica, and silver are assumed to be $-95.92-i10.97$, 2.1025, and $-143.49-i9.52$, respectively. The effective index of the even gap-plasmon mode, calculated using the $+$ sign in \myeqno{pw1it}, is given in \mytabno{tab4}. \myfigno{fig10} shows plots of the LHS and RHS of \myeqno{pw1it} and the convergence of the real and imaginary parts of the effective index. 

The thin 50-nm slab considered above does not support an odd (antisymmetric) gap-plasmon mode. However, increasing the silica thickness to $3\mum$ allows the structure to support gap-plasmon modes of both symmetries. We calculate the effective indices iteratively using \myeqno{pw1it} with $+$ and $-$ signs for odd and even gap-plasmon modes respectively. Once again, the iterative method agrees with the calculations of the \texttt{FindRoot} function in  \mtica{} (\mytabno{tab4}).
\begin{table}[!t]
\centering
	\begin{tabular}{m{1.4cm}|m{1.5cm}|m{3cm}|m{3cm}|m{0.6in}@{}}
	\hline 
	{\bf Mode}&{\bf Equation}&{\bf Iterative scheme}&{\bf \mtica{}}&{\bf $\mathbf{H}$-field}\\  \hline \hline
	Even gap plasmon&\myeqno{pw1it} $\qquad+$&$2.017122399636765 - i0.023755375876767$&$2.017122399636765 - i0.023755375876767$&\includegraphics[width=0.6in]{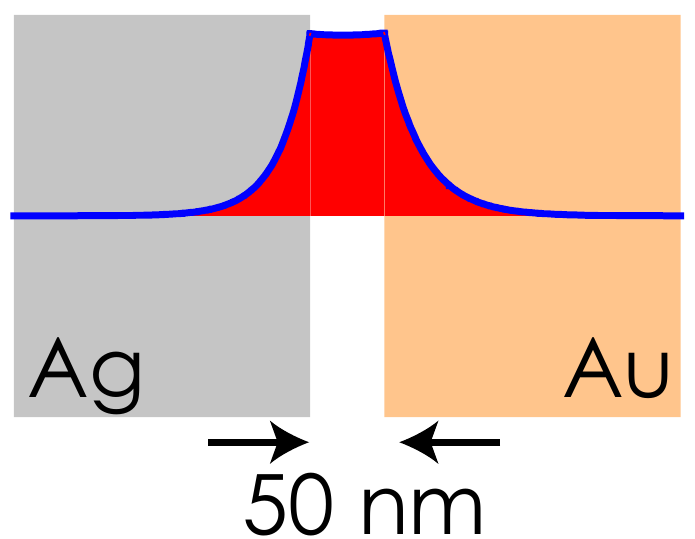}\\ \hline
	 Even gap plasmon&\myeqno{pw1it} $\qquad+$&$1.467915033129527 - i0.001514007231254$&$1.467915033129527 - i0.001514007231254$&\includegraphics[width=0.6in]{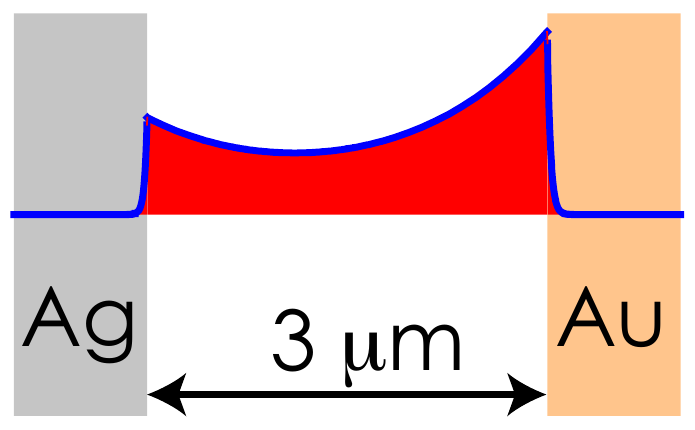}\\ \hline
	 Odd gap plasmon&\myeqno{pw1it} $\qquad-$&$1.455036275034357 - i0.001440093524486$&$1.455036275034357 - i0.001440093524486$&\includegraphics[width=0.6in]{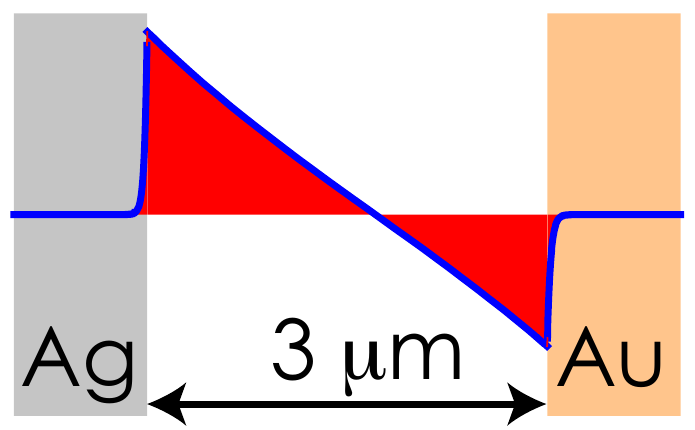}\\ \hline
	 TM$_1$&\myeqno{sw2it} $M=0,-$&$0.007407516660127 - i1.981855964604849$&$0.007407516660127 - i1.981855964604849$&\includegraphics[width=0.6in]{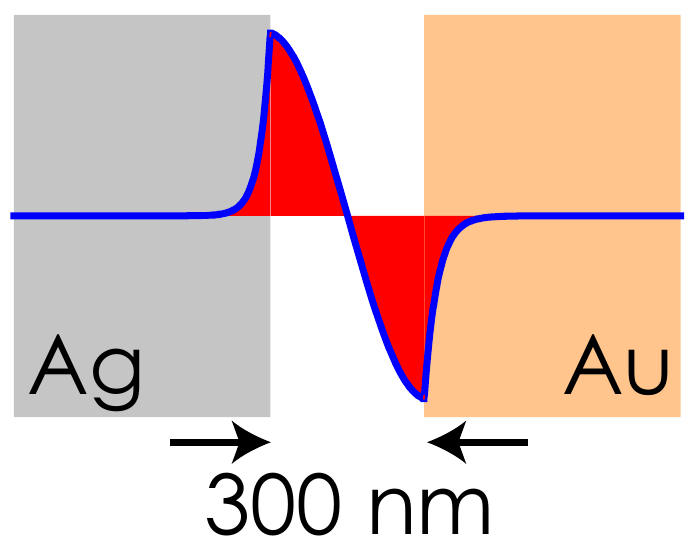}\\ \hline
	 TM$_2$&\myeqno{sw2it} $M=1,+$&$0.001924784371747 - i4.90109582884017$&$0.001924784371747 - i4.90109582884017$&\includegraphics[width=0.6in]{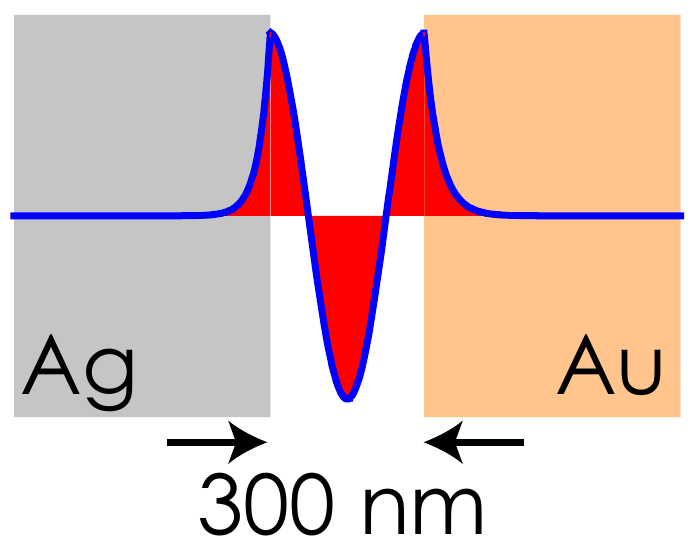}\\ \hline
	 TM$_3$&\myeqno{sw2it} $M=1,-$&$0.000214216445512 + i 7.58348752253199$&$0.000214216445512 + i7.58348752253199$&\includegraphics[width=0.6in]{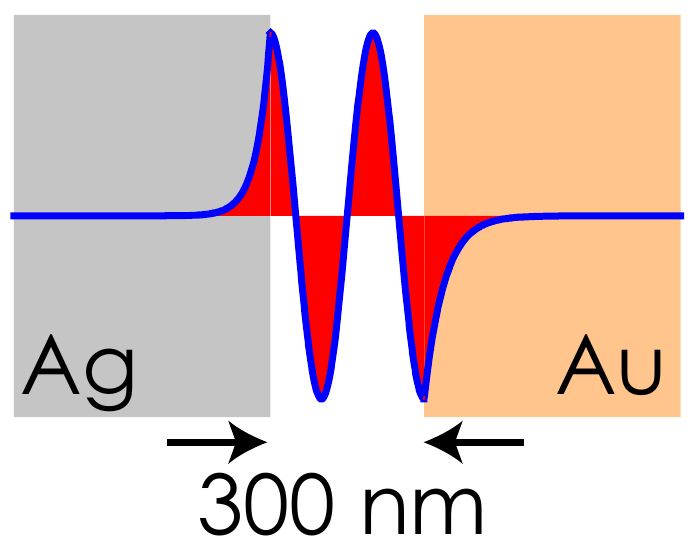}\\ \hline
	TM$_4$&\myeqno{sw2it} $M=2,+$&$0.00592749529203 + i10.22010371292752$&$0.00592749529203 + i10.22010371292752$&\includegraphics[width=0.6in]{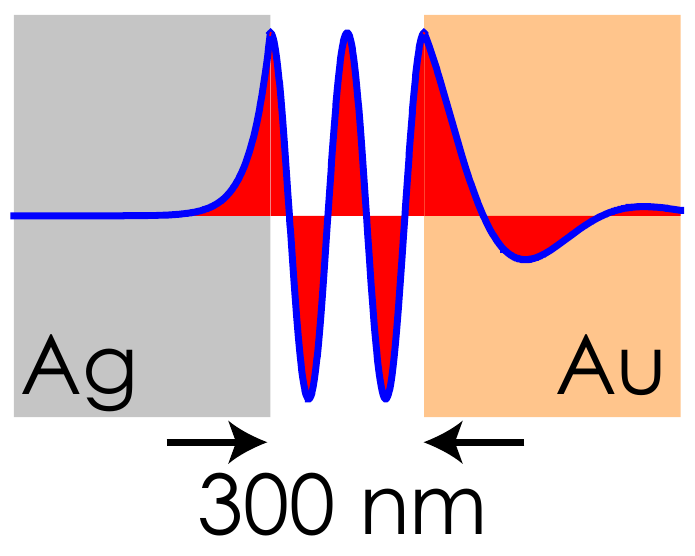}\\ \hline
	TM$_5$&\myeqno{sw2it} $M=2,-$&$0.01577537648440 + i12.83149770403419$&$0.01577537648440 + i12.83149770403419$&\includegraphics[width=0.6in]{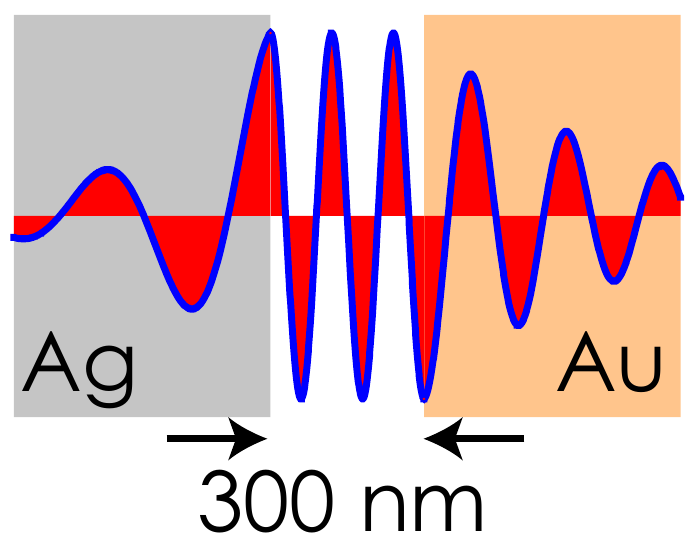}\\ \hline
	\end{tabular} 
\caption{Effective indices of various modes of MDM waveguides obtained using the iterative method, with a comparison to the direct solutions calculated using \mtica{}.}
\label{tab4}
\end{table}
\subsubsection{TM-like waveguide modes}
The other class of modes supported by a MDM waveguide are the TM-like waveguide modes with profiles as shown in \myfigno{fig4}(f) and (g). For these modes,  $\kappa$ is purely imaginary and plasmon \myeqno{pwdisp} transforms to dielectric \myeqno{dwdisp}. The iterative form for obtaining the indices of these modes is identical to \myeqno{sw2it}, with only a few slight differences as regards its implementation. For the case of dielectric waveguides, the mode index $M$ assumed integer values starting from 0 for even modes and 1 for odd modes. For MDM waveguides, this is reversed: $M$ assumes integer values starting from 1 for even modes and 0 for odd modes. This is a consequence of the signs of $p$ and $q$ being reversed for MDM waveguides due to the negative permittivity of the metal ``claddings.'' Additionally, because of the large index difference between metals and dielectrics, the TM-like modes of MDM waveguides are almost always calculated using the strong-confinement formula in \myeqno{sw2it}.

To show how TM-like modes are calculated for MDM waveguides, we choose a 300-nm thick silica layer sandwiched between semi-infinite gold and silver layers. The permittivities of each material are the same as assumed in the previous subsection. The indices of the first five TM-like modes, calculated using the strong-confinement formula \myeqno{sw2it}, are listed in \mytabno{tab4}  alongside the corresponding solutions from \mtica{}. 
\subsubsection{Symmetric MDM waveguides}
Many contemporary high-confinement architectures employ the symmetric MDM waveguide as their skeleton structure. Equality of the relative permittivities of the substrate and cover further simplify \myeqno{pw1it} for the gap-plasmon modes. We can express the resulting iterative forms as:
\begin{subequations}\label{smim}
\begin{align}
\kappa_{n+1}&=-p\sqrt{\kappa_n^2+\Kc^2}\tanh\left(\kappa_n h/2\right)&\text{\ldots for even gap plasmon.}\\
\kappa_{n+1}&=-p\sqrt{\kappa_n^2+\Kc^2}\coth\left(\kappa_n h/2\right)&\text{\ldots for odd gap plasmon.}
\end{align}
\end{subequations}

\subsection{Dielectric-metal-dielectric waveguides}
The second type of plasmonic waveguide structure commonly encountered in practice is an DMD waveguide, as shown in \myfigno{fig4}(d). In theory, metal films of arbitrary thickness in a homogenous dielectric medium (including air) or in Kretschmann-type coupling configurations \cite{Raether1988} are examples of DMD waveguides. In practice, we refer to metallic waveguides as DMD-type only if the SPP modes on the two metal-dielectric interfaces are coupled. Because of the rapid decay of the fields (with distance) inside metals, such a mode-coupling is possible only for thin ($h <100\mnm$) metal films. 

To obtain the iterative form for determining mode indices of DMD waveguides, we write \myeqno{pwdisp} as:
\begin{equation}\label{pw2}
\tanh{\kappa h}=\frac{-2A\kappa}{\kappa^2+A^2-B^2},
\end{equation}
where $A$ and $B$ are defined in \mytabno{tab1}. Considering this as a quadratic equation in $A$ and $B$ leads us to the desired iterative forms. We start by identifying the lower-index dielectric as the substrate and using initial guesses of  $A_0=k_0$, $B_0=0$, and $ \kappa_0=k_0$. We then iterate to obtain five different quantities successively, using the following equations in the order shown:
\vspace{0.1in}
\begin{subequations}\label{pw2it}
\fbox{\parbox[c][1.4in]{5.1in}{\begin{align}
a_n^*&=-\kappa_n\coth{\kappa_nh}\pm\sqrt{B_n^2+\kappa_n^2\csch^2(\kappa_nh)},\label{ait}\\
b_n^*&=\sqrt{a_n^{*2}+\kappa_n^2+2a^*\kappa_n\coth^2(\kappa_nh)},\\
\kappa_{n+1}&=\sqrt{\left(a_n^*+b_n^*\right)^2/p^2+\Qc^2},\\
A_{n+1}&=\left(p\xi_{\mathrm{c,}n+1}+q\xi_{\mathrm{s,}n+1}\right)/2,\\
B_{n+1}&=\left(p\xi_{\mathrm{c,}n+1}-q\xi_{\mathrm{s,}n+1}\right)/2.
\end{align}}}\vspace{0.1in}
\end{subequations}

Here, $a_n^*, b_n^*$ are intermediate dummy parameters and $\xi_{\mathrm{c,}n+1}$ and $\xi_{\mathrm{s,}n+1}$ are related to $\kappa_{n+1}$ through \mytabno{tab1}. The indices of the low- and high-energy plasmon modes are obtained by using the $+$ and the $-$ signs, respectively, in \myeqno{ait}. The large asymmetry in the decay constants in the metal core ($\kappa$) and dielectric claddings ($\xc,\xs$) makes the determination of the effective indices of DMD modes a numerically-challenging problem. For the iterative procedure, this translates into a difficulty in obtaining convergent iteration expressions. Therefore, unlike the previous cases, the iterative procedure for DMD waveguides involves multiple steps instead of a single closed-form iteration function. Admittedly, this multi-step algorithm goes against the ``pocket calculator'' philosophy, but nevertheless is convenient and efficient compared to direct numerical solutions.
	\begin{figure}[!tb]
	\centering
	\includegraphics[width=13cm]{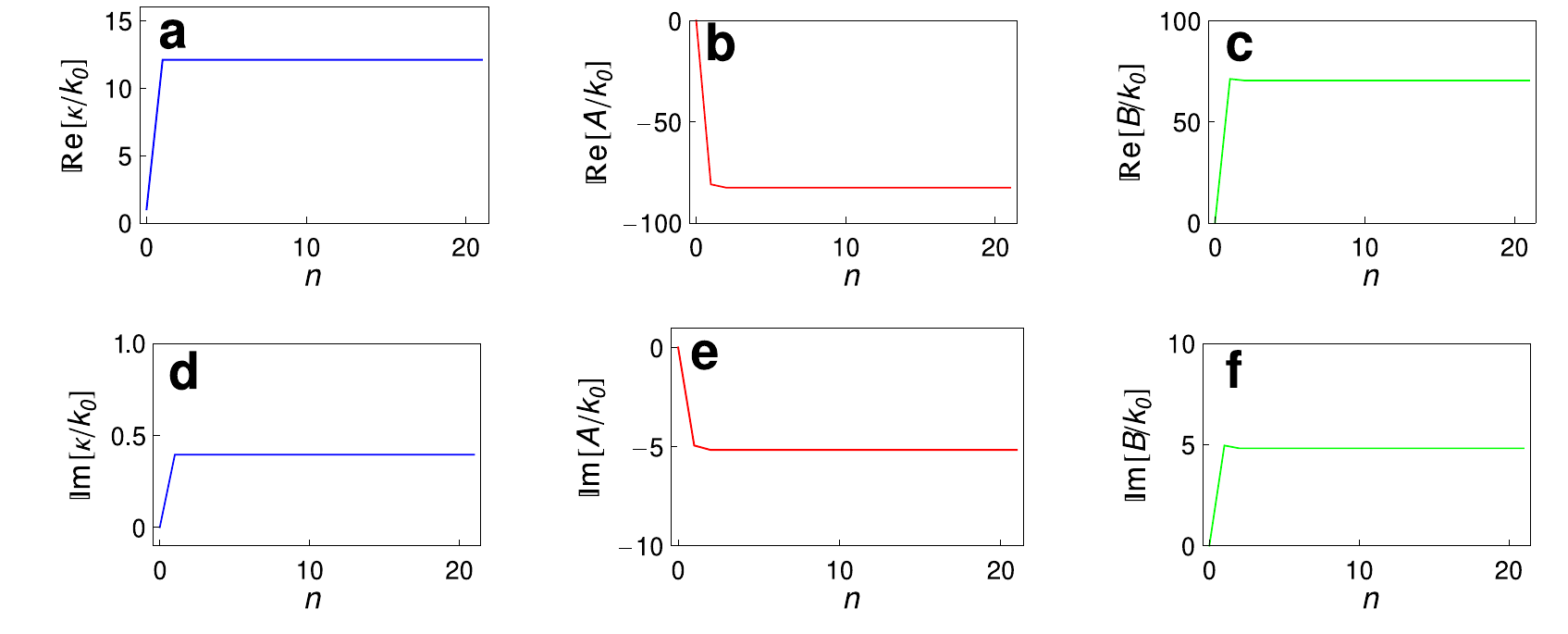}
	\caption{Convergence of the real and imaginary parts of $\kappa$, $A$, and $B$ in \myeqno{pw2it} for the case of a 50 nm thick silver-silica-silver waveguide operating at 1550 nm.}
	\label{fig11}
	\end{figure}
\begin{table}[!bt]
\centering
	\begin{tabular}{m{1cm}|m{1.3cm}|m{2.8cm}|m{2.9cm}|m{0.6in}@{}}\hline 
	{\bf Mode}&{\bf Equation}&{\bf Iterative scheme}&{\bf \mtica{}}&{\bf $\mathbf{H}$-field}\\  \hline \hline
	High energy mode&\myeqno{pw2it} $\qquad-$&$1.4610633883905-i0.0008056177064$&$1.46106338839051-i0.0008056177064$&\includegraphics[width=0.6in]{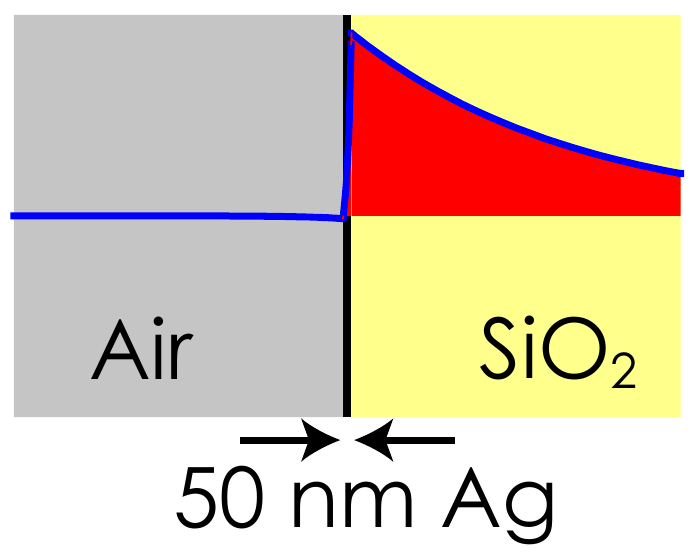}\\ \hline
	 Low energy mode&\myeqno{pw2it} $\qquad+$&$1.4603904174862-0.0006470130493$&$1.46039041748617-i0.0006470130493$&\includegraphics[width=0.6in]{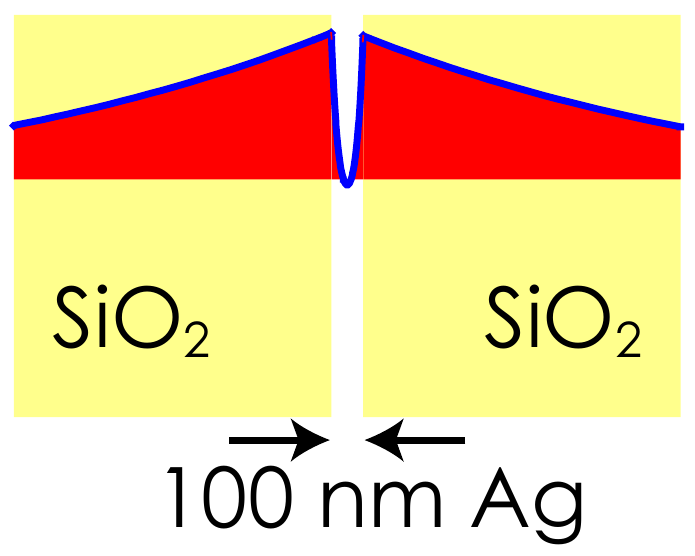}\\ \hline
	 High energy mode&\myeqno{pw2it} $\qquad-$&$1.4610140056811-i0.0007906968233$&$1.4610140056808-i0.0007906968233$&\includegraphics[width=0.6in]{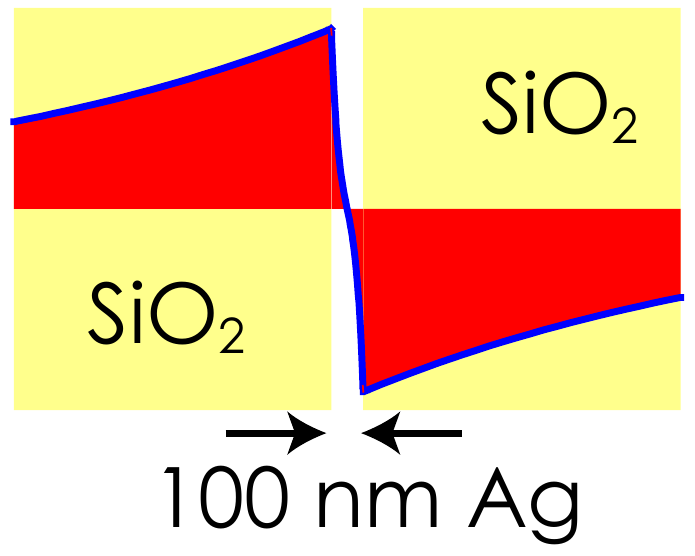}\\ \hline
	\end{tabular} 
\caption{Effective indices of various modes of an DMD waveguide obtained using the iterative method and their comparison with direct solutions using \mtica{}.}
\label{tab5}
\end{table}

We will now illustrate the use of \myeqno{pw2it} through specific examples. Consider the operation of a 50-nm-thick silver film on a silica substrate at 1550 nm. The relative permittivities of silver and silica are the same as assumed previously; the top cover is air ($\epsc=1$). We start by specifying $A_0\,,B_0$, and $\kappa_0$, and obtain their successive values according to \myeqno{pw2it}. The convergence of $A_n\,,B_n$, and $\kappa_n$ is shown in \myfigno{fig11}; \mytabno{tab5} compares the solution obtained using the iterative scheme with the direct solution computed by the \mtica{} \texttt{FindRoot} function. In our computations, the \texttt{FindRoot} function was unable to return the complex mode index unless we gave a very precise initial guess for both the real \emph{and} the imaginary parts. On the other hand, the iterative method computed the complex mode index regardless of the initial guess.
 
Our next example is a symmetric structure with a 100-nm silver film sandwiched between two silica layers. Although 100 nm is at the boundary of the film behaving like an DMD waveguide versus two separate interfaces, we choose these dimensions to highlight the robustness of the iterative technique even for the most extreme cases of root-finding. For this structure, both high- and low-energy modes exist, whose indices are conveniently calculated by using the $-$ and $+$ signs, respectively, in \myeqno{pw2it}. The indices calculated using the iterative method and \mtica{} for this example are included in \mytabno{tab5} and show good agreement. Calculating these indices was especially difficult using \texttt{FindRoot} because of their close numerical proximity. The iterative technique, on the other hand, specifies separate functions (the $+$ and$-$ signs) which are guaranteed to converge to these different modes, irrespective of their numerical proximity.  
\subsubsection{Symmetric DMD waveguides}
Symmetric DMD waveguides appear in the form of idealized waveguide geometries such as metal films in homogenous dielectric media (including air). Fortunately, for the symmetric  case, the iterative scheme consists of a single equation which can be written as:
\begin{subequations}\label{simi}
\begin{align}
\kappa_{n+1}&=\frac{\Qc}{\sqrt{1-p^{-2}\tanh^2\left(\kappa_n h/2\right)}}&\text{\ldots for even plasmon mode.}\\
\kappa_{n+1}&=\frac{\Qc}{\sqrt{1-p^{-2}\coth^2\left(\kappa_n h/2\right)}}&\text{\ldots for odd plasmon mode.}
\end{align}
\end{subequations}
These equations do not follow automatically from \myeqno{pw2it} but instead have to be derived separately by considering the dispersion equation for the symmetric DMD waveguide. \myeqno{simi} are preferable to \myeqno{pw2it} for analyzing symmetric DMD structures owing to better convergence behavior and an obvious ease in programming. 

\section{Conclusion}
We have presented a robust and an easy-to-implement iterative technique for determining complex propagation constants of asymmetric dielectric and plasmonic waveguides. At the heart of our procedure are the closed-form iteration functions, namely the boxed Eqs.~\eqref{sw2it}, \eqref{sw3it}, \eqref{pw1it}, and \eqref{pw2it}. In addition to the programming ease, the iterative technique has an inherent ability to give arbitrary-precision answers---a feat difficult to achieve using graphical or curve-fitting algorithms such as the reflection-pole method. Because of its insensitivity to initial guess, it is our hope that this technique will help facilitate design tasks that require rapid and automated calculation of mode indices for a variety of nanophotonic structures having a rectangular geometry.
\section*{Acknowledgements}
We thank Dr. E. Kocaba\c{s} and Prof. D. A. B. Miller for several helpful discussions. We also gratefully acknowledge the financial support from the Si-based Laser Initiative of the Multidisciplinary University Research Initiative (MURI) under the Air Force Aerospace Research Award No. FA9550-06-1-0470 and the MARCO Interconnect Focus Center.
\end{document}